\documentclass[sigplan,10pt]{acmart}

\renewcommand\footnotetextcopyrightpermission[1]{}

\usepackage{tikz}
\usepackage{float}
\usepackage{xcolor}
\usepackage{subfig}
\usepackage{amsmath}
\usepackage{physics}
\usepackage{booktabs}
\usepackage{colortbl}
\usepackage{enumitem}
\usepackage{graphicx}
\usepackage{textcomp}
\usepackage{algorithm}
\usepackage{tcolorbox}
\usepackage{algpseudocode}
\definecolor{blue_color}{rgb}{0.0, 0.0, 1.0}
\definecolor{red_color}{rgb}{1.0, 0.0, 0.0}

\newcommand*\circled[1]{\tikz[baseline=(char.base)]{
            \node[shape=circle,fill,inner sep=1pt] (char) {\textcolor{white}{#1}};}}
          
\newcommand{\sol}{\textsc{Spyce}}
\newcommand{\cx}{\texttt{CX}}
\newcommand{\sx}{\texttt{SX}}
\newcommand{\rz}{\texttt{RZ}}
\newcommand{\rx}{\texttt{RX}}
\newcommand{\x}{\texttt{X}}
  
\begin{document}
\sloppy

\title{Toward Privacy in Quantum Program Execution On Untrusted Quantum Cloud Computing Machines for Business-sensitive Quantum Needs}

\author {
    \text{Tirthak Patel$^\dag$,
    Daniel Silver$^\S$,
    Aditya Ranjan$^\S$,
    Harshitta Gandhi$^\S$,
    William Cutler$^\S$,
    Devesh Tiwari$^\S$}\newline
    \textit{\newline $^\dag$Rice University} \; \textit{$^\S$Northeastern University}
}

\begin{abstract}

Quantum computing is an emerging paradigm that has shown great promise in accelerating large-scale scientific, optimization, and machine-learning workloads. With most quantum computing solutions being offered over the cloud, it has become imperative to protect confidential and proprietary quantum code from being accessed by untrusted and/or adversarial agents. In response to this challenge, we propose \sol{}, which is the first known solution to obfuscate quantum code and output to prevent the leaking of any confidential information over the cloud. \sol{} implements a lightweight, scalable, and effective solution based on the unique principles of quantum computing to achieve this task.

\end{abstract}

\maketitle
\pagestyle{plain}

\section{Introduction to \sol{}}
\label{sec:introduction}

Quantum computing is an emerging technology that has the potential to accelerate and make possible the execution of many large-scale scientific, optimization, and machine-learning tasks~\cite{preskill2021quantum,bova2021commercial}. As quantum computing technology advances, multiple cloud-based quantum computing platforms are being used to develop and execute classically-infeasible mission-critical tasks by government agencies and industry partners~\cite{raymer2019us,gibney2017billion,gibney2019quantum}. In many cases, the solutions to these tasks are business sensitive and should be protected (e.g., the solution to a classically-infeasible problem relevant to a defense program). Currently, due to the nascent stage of quantum cloud computing, the cloud computing providers have full access to the end users' mission-sensitive programs and the output of such programs~\cite{saki2021survey,phalak2021quantum}. 

Recognizing the importance of security and privacy for quantum program execution, there has been some related work on it, although not solving the same problem as this work (protecting the output of quantum programs). In particular, encrypting quantum information over networks~\cite{xu2023nested,zhang2022instantiation,balaji2022reliable} and securing quantum programs from third-party quantum compilers~\cite{suresh2021short,saki2021split} have received attention.

Unfortunately, all of these works assume that the cloud hardware provider is an uncompromised entity and does not have intentional or unintentional snoopers on the quantum cloud platform that can analyze the program outputs. Even if the code is protected from the compiler and over the network~\cite{xu2023nested,zhang2022instantiation,balaji2022reliable,suresh2021short,saki2021split}, currently, it has to be decrypted before it can be run on the hardware so that the correct output can be obtained, which is open to snooping from the cloud provider. Even if the cloud provider is uncompromised, organizations may not want to disclose their tasks, proprietary code, and program solutions to the cloud provider. Protecting this information from the cloud provider is a non-trivial challenge as \textit{the user essentially wants the hardware provider to run the ``wrong'' code and observe the ``wrong'' output, but be able to recover the ``correct'' quantum output from the ``wrong'' output on the user's end. We propose \sol{} to achieve just this.} 

In the near future, it is anticipated only a few entities in the world may have access to powerful quantum computers, and these quantum computers will be used to solve previously-unsolved large-scale optimization problems, possibly without an explicit trust model between the service cloud provider and the customer. Therefore, the solutions to such large-scale optimization problems will be considered sensitive and will need to be protected. \sol{} takes the first few steps toward preparing us for that future -- by developing a novel method that intelligently obfuscates program output and quantum circuit structure of the original quantum program provided by the user/customer. 

Before we introduce the contributions of \sol{}, we first provide a primer on relevant quantum computing concepts.

\begin{figure}[t]
    \centering
    \includegraphics[scale=0.47]{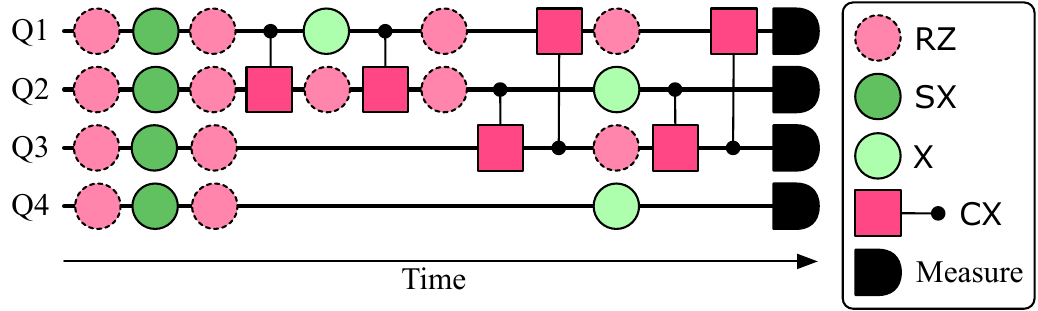}
    \vspace{1mm}
    \hrule
    \vspace{-3mm}
    \caption{Example circuit representation of a quantum algorithm. The horizontal lines represent qubits with gates being applied to them in order from left to right.}
    \vspace{-5mm}
    \label{fig:base_circuit}
\end{figure}

\vspace{2mm}

\noindent\textbf{Qubits and Quantum States.} The fundamental unit of quantum computing is the \textit{qubit}, which is capable of representing a \textit{superposition} (linear combination) of two orthogonal basis states. This is represented as $\ket{\Psi} = \alpha\ket{0} + \beta\ket{1}$, where $\alpha$ and $\beta$ are the complex amplitudes of the constituent basis states. Upon measurement, this superposition collapses such that the probability of measuring the state $\ket{0}$ is $\norm{\alpha}^2$ and $\norm{\beta}^2$ for measuring the $\ket{1}$ state, where $\norm{\alpha}^2 + \norm{\beta}^2 = 1$.

A general system of $n$ entangled qubits is represented as a linear combination (superposition) of $2^n$ basis states: \mbox{$\ket{\psi} = \sum_{k=0}^{k=2^n-1} \alpha_k\ket{k}$}. As with one qubit, when the multi-qubit state is measured, it manifests as a projection to one of the basis states (where now the basis state is a state of $n$ qubits). The probability of measuring state $\ket{k}$ is $\norm{\alpha_k}^2$.

\vspace{2mm}

\noindent\textbf{Quantum Gates and Circuits.} Quantum gates are applied to manipulate the qubit state. The universal basis gate set for IBM's quantum computers (used in this work) consists of the \cx{}, \sx{}, \x{}, and \rz{} gates~\cite{ibmquantum}. \sx{}, \x{}, and \rz{} gates are one-qubit gates. In the Bloch-sphere representation, where a qubit's superposition state is represented on the surface of a sphere, one-qubit gates are categorized by the axis around which the rotation takes place and the angle of rotation. The generalized rotation gate \rz{}$(\theta)$ rotates the qubit state about the z-axis by angle $\theta$. Similarly, the generalized \rx{}$(\beta)$ gate performs a rotation about the x-axis by angle $\beta$. The generalized \rx{} gate is decomposed into the basis gates before execution on the IBM platform. Note that \x{}$=$\rx{}$(\pi)$ and \sx{}$=$\rx{}$(\frac{\pi}{2})$.

The \cx{} (controlled-\x{}) gate is a two-qubit gate that enables entanglement and applies the \x{} gate to the ``target'' qubit only when the ``control'' qubit is $\ket{1}$. As a note, all of these gates can be represented using unitary matrices. A unitary matrix is a matrix $U$ such that $U^\dagger U = I$, where $U^\dagger$ is the complex conjugate transpose of $U$ and $I$ is the identity matrix.

A quantum program comprises a sequence of one- and multi-qubit gates, as shown in Fig.~\ref{fig:base_circuit}. The sequence is referred to as a \textit{quantum circuit}. A quantum circuit also has a unitary matrix representation, $U$, which can be calculated by taking a Kronecker product of the unitary matrices of all the gates in the circuit in sequential order. At the end of the sequence, the qubits are measured to get the output (answer to the problem being solved). This circuit is required to be prepared and measured multiple times to get a probability distribution over its basis states. \textit{This probability distribution is the output of the quantum program, referred to as program output (solution).}

\begin{figure}[t]
    \centering
    \includegraphics[scale=0.43]{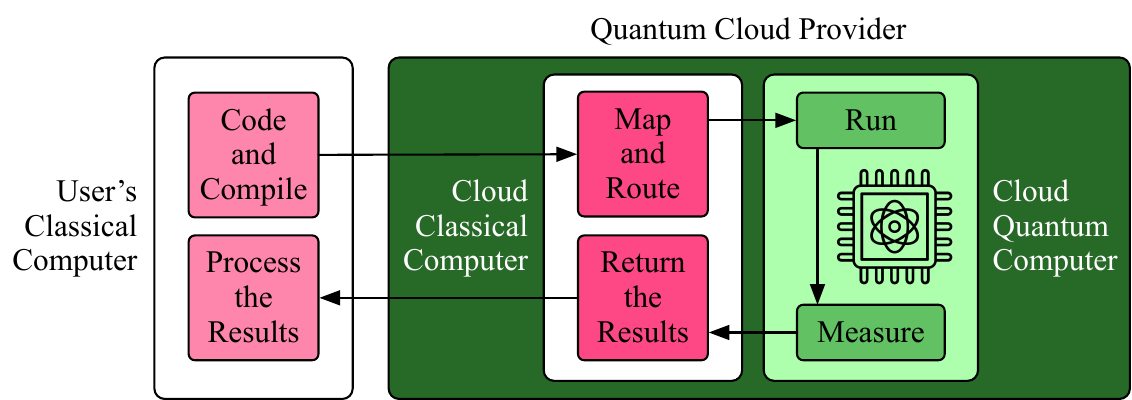}
    \vspace{1mm}
    \hrule
    \vspace{-3mm}
    \caption{A quantum algorithm's execution workflow consists of several steps on the user's side as well as the cloud.}
    \vspace{-5mm}
    \label{fig:workflow}
\end{figure}

\vspace{2mm}

\noindent\textbf{Quantum Computing Hardware and Noise.} Currently, there are multiple quantum technologies in different stages of development including superconducting qubits, ion traps, neutral atoms, and photons. While \sol{} is generally applicable to any type of technology, in this paper, we evaluate it on IBM's superconducting-qubit computers as they have the advantage of being relatively easy to fabricate and operate, which has made them a popular choice for quantum computing research~\cite{li2022paulihedral,patel2022quest,balaji2022reliable}. These computers implement a qubit using a Josephson junction, where two superconductors are separated by a thin insulating layer to create a tuneable potential energy barrier for the $\ket{0}$ and $\ket{1}$ states~\cite{patel2020disq}.

A challenge in executing programs on real quantum computers is hardware noise effects. These noise effects include the \textit{state preparation and measurement (SPAM) errors}, which cause the qubit to be incorrectly initialized and measured, the \textit{gate errors}, which refer to the qubit state being incorrectly modified, and the \textit{decoherence errors}, which refer to the loss of the coherence of a qubit's state. These errors are an order of magnitude higher for the \cx{} gates than the \x{} and \sx{} gates~\cite{patel2020experimental}; \rz{} gates do not have any error as they are implemented virtually using change of reference. The impact of hardware noise on program output adds an additional challenge for \sol{} as it cannot apply obfuscation techniques that would increase the effects of hardware noise; the quality of the recovered ``correct'' program output should not be worse than what the user would have observed if the output was not obfuscated. \textit{A useful quantum computing system software solution must demonstrate its effectiveness on current noise-prone quantum computers -- this is why \sol{} is designed for and evaluated on real quantum hardware.} 

\vspace{2mm}

\noindent\textbf{Quantum Execution Workflow.} Fig.~\ref{fig:workflow} shows the execution workflow of a quantum computing workload. The user writes the code and transpiles the circuit on their end. Then the code is sent to the cloud, where the mapping and routing operations are performed to convert the circuit to a format that can be executed on the qubit connectivity of the selected quantum computer. Note that these steps can also be performed on the user's end, in which case the hardware provider simply runs the circuit. \textit{Where exactly these steps are performed is not of consequence to \sol{}, as \sol{} is applied before these steps are run.} Then, the compiled circuit is run on the hardware, and the output is measured and returned to the user, who processes it for analysis.

\vspace{2mm}

\noindent\underline{\textbf{Limitations of Existing Work.}} Considering the above execution workflow, most previous works related to executing quantum codes privately in the quantum cloud, referred to as ``blind quantum computation''~\cite{fitzsimons2017private,zhang2021succinct,xu2021verification}, are not applicable to the NISQ era as they assume that the user has access to at least some local quantum computational resources and the ability to transmit these qubit states over a quantum internet to the quantum cloud for computation. In terms of the current NISQ era, previous work has been limited to the network and compiler domain, as described below. While useful, these approaches do not address the problem solved by \sol{}, and cannot be modified to achieve \sol{}'s goals. 

\vspace{2mm}

\noindent\textbf{Quantum Encryption for Communication Networks.} Some previous works have focused on private quantum network protocols~\cite{xu2023nested,zhang2022instantiation,balaji2022reliable}. As an instance, Balaji et al.~\cite{balaji2022reliable} implement a lattice cryptographic technique for post-quantum encryption with robustness against the Man-In-The-Middle and Sybil attacks. By the same token, Xu et al.~\cite{xu2023nested} propose a system to solve the issue of hiding preambles via random repetition and nested hash coding to encrypt network data in a post-quantum manner. In contrast, Zhang et al.~\cite{zhang2022instantiation} use obfuscation over the network as a form of encryption. The proposed approach utilizes quantum point obfuscation by constructing quantum hash functions, which help enhance the privacy of the ciphered data. \textit{These works require that the cloud provider is trusted as the code/data transferred over the network needs to be decrypted before execution.}

\vspace{2mm}

\noindent\textbf{Protection from Third-Party Compiler Software.} Some previous works have also focused on protecting quantum code from third-party compilers~\cite{suresh2021short,saki2021split,saki2021survey,phalak2021quantum}. For example, Suresh et al. ~\cite{suresh2021short} attempt to confuse the compiler by adding \cx{} gates, which are removed after compilation and before hardware execution to get the correct output. Consistent with the above paper, Saki et al.~\cite{saki2021split} propose a split compilation approach where different sections of the code are sent to different compilers so that no one compiler has access to the whole code. The code is stitched back together before hardware execution. On the other hand, Phalak et al.~\cite{phalak2021quantum} propose verification techniques to validate that third-party vendors are indeed dedicating the promised resources to the quantum code. \textit{These works also do not obfuscate the output from the cloud provider as the circuits are decoded pre-execution.}

\begin{figure}[t]
    \centering
    \includegraphics[scale=0.49]{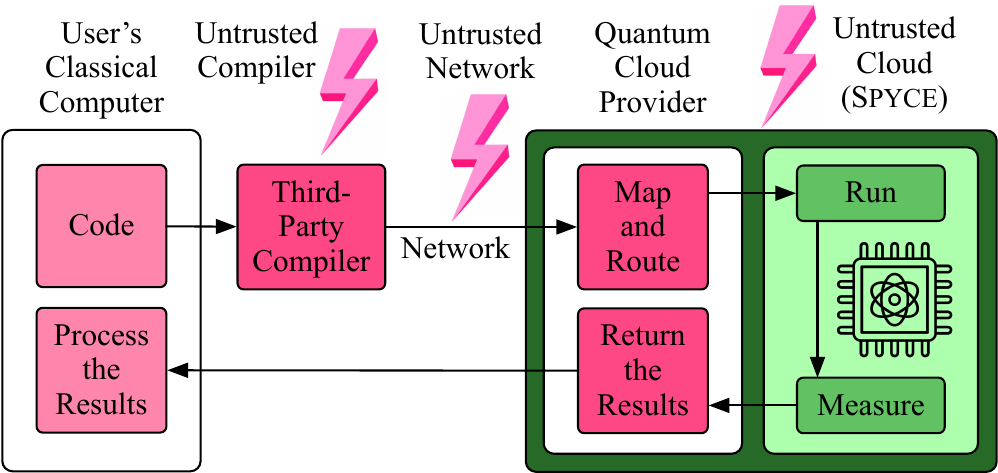}
    \vspace{1mm}
    \hrule
    \vspace{-3mm}
    \caption{Previous works in quantum privacy have either assumed an untrusted compiler or network model. \textit{\sol{} is the first work to also protect code from an untrusted cloud.}}
    \vspace{-4mm}
    \label{fig:related_work}
\end{figure}

\vspace{2mm}

\noindent\underline{\textbf{\sol{}: Scope and Threat Model.}} All prior works assume that quantum cloud hardware providers are trusted and are not snooping on the solutions to the quantum programs they execute on their hardware. In contrast, \sol{} assumes that only the user's local system is trusted. All other components involved in the execution of quantum code including third-party compilers, networks, quantum clouds, and hardware are assumed to be open to snooping. 

We underscore that, although \sol{} obfuscates a quantum program's circuit structure as a side-benefit to make the obfuscation of the circuit output more effective, \sol{} does not make any theoretical claims about privatizing or hiding parts of the original quantum algorithm. Empirically, we demonstrate that \sol{} significantly obfuscates a quantum program's circuit structure using graph-based distances (Sec.~\ref{sec:evaluation}), although this by itself is not \sol{}'s primary goal. 

We note that, while tempting, classical techniques for code obfuscation are not suitable to be applied in the quantum computing domain because of differences in computing models and observability. Quantum computing fundamentally relies on entanglement and superposition principles to perform computation, therefore any injection of randomness or obfuscation cannot be ``disentangled'' without destroying the quantum state and collapsing the computation, as classical randomization or obfuscation technique would attempt to.   

\vspace{2mm}

\noindent\underline{\textbf{\sol{}: Approach and Contributions.}} We propose \sol{} as a simple yet effective obfuscation technique to obfuscate the quantum program output (and its circuit structure) of quantum programs in the quantum cloud. 

\sol{}'s simplicity lies in its approach to randomly \textbf{injecting \x{} gates} (similar to classical \texttt{NOT} gates) at the end of the quantum circuit. The location of \x{} gate injection becomes the key -- there are $2^n$ different permutations of \x{}-gate injections, where $n$ is the number of qubits in the quantum program. This approach allows the program output to become scrambled so that it can only be decoded by the end user who holds the decoding key, but not the cloud provider. 

While promising, as our evaluation demonstrates (Sec.~\ref{sec:evaluation}), only injecting \x{} gates does not provide significant structural divergence despite leveraging quantum circuit synthesis to ``pull'' back the appended \x{} gates and merge them in the inner layers of the circuit. To mitigate this challenge, \sol{} demonstrates how to intelligently inject \rx{} pairs and design a novel quantum circuit synthesis pass that strategically decomposes the new ``\x{} gates and \rx{} pairs added'' circuit into multiple blocks, transforms them in semantically equivalent but structurally different blocks, and then, combines them together -- the resultant circuit, by design, successfully obfuscates the location of added \x{} gate and \rx{} gate. \sol{}'s design demonstrates: (1) how to exploit the reversibility property of quantum gates to ensure that injected gates do not alter the original program logic and yet obfuscate the output, (2) how to design its gate injection and circuit synthesis procedure such that the resulting obfuscated circuit does not become more sensitive to quantum hardware errors due to addition of new gates (Sec~\ref{sec:design}). Obfuscating circuit output and structure is naturally likely to increase gate count and depth (as a side effect) -- but, \sol{} demonstrates how its impact can be limited on real quantum hardware via a careful design. 

\begin{figure*}
    \centering
    \includegraphics[scale=0.42]{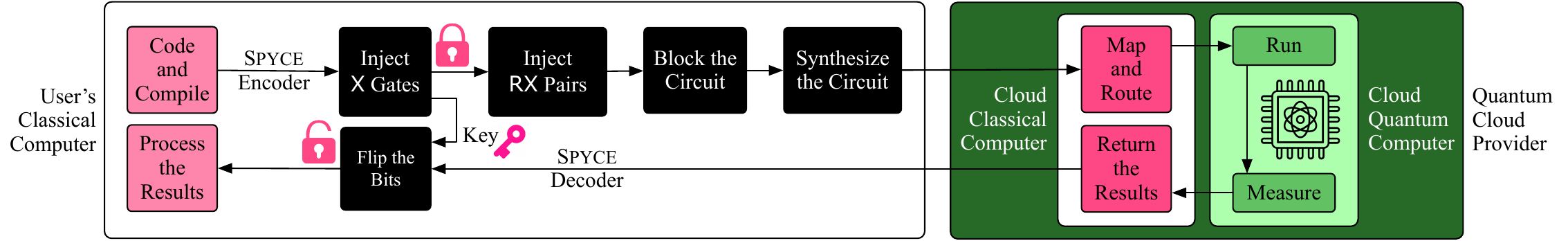}
    \vspace{1mm}
    \hrule
    \vspace{-3mm}
    \caption{\sol{}'s encoding is a four-step and decoding is a one-step process. \sol{}'s steps run entirely on the user/client side.}
    \vspace{-3mm}
    \label{fig:spire_workflow}
\end{figure*}

\vspace{2mm}

\noindent{\textbf{The contributions of this work are as follows.}

\vspace{-1mm}

\begin{itemize}[leftmargin=*]

    \item \sol{} presents the first approach to obfuscate quantum program circuits and outputs in quantum cloud environments -- to ensure that quantum cloud providers cannot infer the solutions to previously unsolved, classically infeasible, large-scale optimization problems which can be business-sensitive and need to be protected.
    
    \vspace{1mm}

    \item \sol{} presents a design and implementation of a simple yet effective framework for achieving its goal by an intelligent combination of \x{} gate and \rx{} gate injections and a novel quantum circuit synthesis procedure.  \sol{} obfuscates the circuit output successfully, and despite the injection of additional quantum gates, \sol{} maintains the same solution quality as the original un-obfuscated quantum circuit -- in the presence of quantum hardware errors.  

    \vspace{1mm}

    \item Our extensive simulation and real-hardware evaluation of a diverse set of algorithms (up to 128 qubits) demonstrates that \sol{} \textit{successfully hides the program output, solution states, and circuit structures}, using relevant probability distribution distance and circuit structure distance metrics, while maintaining low compilation times and output error. 

    \vspace{1mm}

    \item The paper also presents an in-depth analysis of applying \sol{} to a widely-used quantum iterative optimization algorithm, QAOA~\cite{farhi2016quantum}, for finding a solution to the MaxCut problem. \sol{} demonstrates that it is not possible to solve this optimization problem if one only relies on the code and scrambled output that the cloud would observe. In contrast, \sol{} has similar solution quality as the un-obfuscated version without increasing the optimization time.

\end{itemize}

\section{\sol{}: Design and Implementation}
\label{sec:design}

We begin this section by providing an overview of \sol{}'s design and then delve into the design details of each step. 

\subsection{Overview of \sol{}'s Design}

Fig.~\ref{fig:spire_workflow} shows the execution workflow of a quantum program with \sol{}'s steps included. \sol{}'s encoding process consists of four steps. First, to hide the output, \sol{} randomly selects qubits and injects \x{} gates at the end of the quantum circuits. This gives $2^n$ different permutations of \x{}-gate injections, where $n$ is the number of qubits in the quantum program. The chosen permutation becomes the decoding key when the output is returned after circuit execution. Second, to obfuscate the structure of the circuit, \sol{} injects \rx{} gates with random rotations throughout the code.

As the third and fourth steps, to obfuscate the location of where the \rx{} and \x{} gates are inserted, \sol{} divides the circuit into two-qubit blocks and synthesizes them into a different gate structure that is mathematically equivalent to the original logic. At the end of these four steps, \sol{}-generated code has a significantly different circuit structure and program output distribution compared to the original code -- this \sol{}-generated code is sent to be executed on the quantum cloud. \sol{}'s decoder is a simple, one-step procedure. Post execution, when the program output is returned to the user, the bits in the output state need to be flipped in accordance with the key generated during the encoding process. Once this is done, the correct output is obtained, which only the user has access to.

Next, we describe the \x{}-gate injection step.

\subsection{Output Obfuscation using \x{}-Gate Injections}

Injecting \x{} gates at the end of the circuit (after all other gates in the original circuit logic have been executed) performs the function of flipping the output state bits. For example, consider a two-qubit program output with the following output probability distribution: $p(\ket{00}) = 0.6$, $p(\ket{01}) = 0.1$, $p(\ket{10}) = 0.1$, and $p(\ket{11}) = 0.2$. Let us set our key to be $01$, i.e., we insert an \x{} gate on and flip the first qubit (the least significant digit) and we do not insert an \x{} gate on the second qubit (the most significant digit). This key will perform the following transformation: $\ket{00} \rightarrow \ket{01}$, $\ket{01} \rightarrow \ket{00}$, $\ket{10} \rightarrow \ket{11}$, and $\ket{11} \rightarrow \ket{10}$. Post the \x{}-gate injection, the output probability distribution becomes: $p(\ket{00}) = 0.1$, $p(\ket{01}) = 0.6$, $p(\ket{10}) = 0.2$, and $p(\ket{11}) = 0.1$. This output probability distribution is different from the output distribution of the original distribution. In fact, the highest probability state, which is usually the most important state for most quantum algorithms, shifts from $\ket{00}$ to $\ket{01}$. Thus, it cannot be identified correctly post-obfuscation. \textit{\sol{} specifically chooses \x{} gates for injection because \x{} gates serve as pure inverters, which makes it possible to decode the output with the key. Injecting any other quantum gate with arbitrary angles would require high-overhead and impractical state tomography procedures to deduce the original program output. We discuss some of these other procedures that we considered in Appendix A (see supplementary materials).}

\begin{figure*}[t]
    \centering
    \includegraphics[scale=0.45]{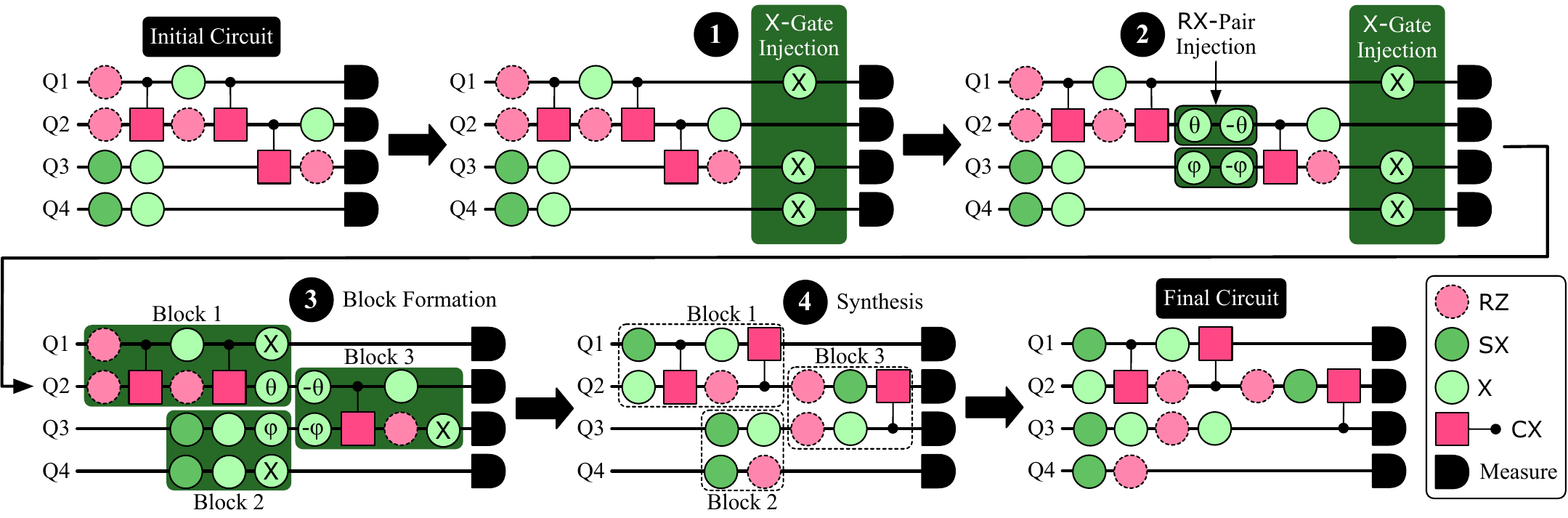}
    \vspace{1mm}
    \hrule
    \vspace{-3mm}
    \caption{\sol{}'s four-step encoding procedure. (1) Inject \x{} gates on arbitrary qubits to obfuscate the output. (2) Inject \rx{} pairs with random rotation angles throughout the circuit to obstruct the circuit structure. (3) Divide the circuit into several two-qubit blocks. (4) Synthesize the blocks into new gate logic, generating a circuit with obfuscated structure and output.}
    \vspace{-3mm}
    \label{fig:overview}
\end{figure*}

The \x{}-gate injection is shown visually in Fig.~\ref{fig:overview} (Step \circled{1}). While the injected \x{} gates are operationally the same as the other \x{} gates in the circuit, we label them with a ``\x{}'' for visualization purposes. In the above two-qubit example, the key is a randomly selected permutation from a uniform distribution of four possible permutations. Thus, it may appear that it can be guessed. However, in general, for an $n$-qubit output, the key is selected from $2^n$ possible permutations. Thus, it quickly becomes untenable to guess the key for realistic quantum algorithms. For example, a 32-qubit algorithm has over one billion different permutations and a 128-qubit algorithm has over one hundred trillion trillion trillion different permutations. We study algorithms of both of these sizes in the evaluation section (Sec.~\ref{sec:evaluation}). However, while it is not possible to surmise this key for realistic medium-to-large quantum algorithms in a brute-force manner, an adversary may still be able to inspect the circuit structure and identify where the \x{} gates are injected as they have access to the circuit. We will see later in this section how \sol{} hides the injected \x{} gates. Before that, we go over how \sol{} obfuscates the entire circuit structure using \rx{}-gate injections.

\subsection{Structure Obfuscation using \rx{}-Gate Injections}

While the \x{} gates obfuscate the output, the rest of the circuit structure can still be inspected by the adversary to extract the locations of \x{} gate injection. Using circuit synthesis, one can ``pull'' \x{} gates injected at the end of the circuits in the inner layers of the circuits to achieve partial structural obfuscation. While useful, our evaluation (Sec.~\ref{sec:evaluation}) shows that such an approach is not as effective as desired -- that is, it does not yield sufficient structural obfuscation. 

Additionally, structural obfuscation for quantum circuits is useful for other purposes too -- for example, it  prevents the adversary from learning about the functionality of the circuit. It is also useful to obfuscate sub-regions within a quantum circuit as those sub-regions may consist of full quantum algorithms by themselves. As an instance, the Quantum Fourier Transform (QFT) ~\cite{namias1980fractional} logic frequently shows up in other quantum algorithm circuits. Thus, an adversary might be able to identify the region of the quantum circuit that forms the QFT logic if that region is not obfuscated. As another example, \textit{the inputs to quantum algorithms are always supplied as circuit gates (e.g., parameterized rotation angles).} Therefore, it is essential to obfuscate the circuit structure to obfuscate any inputs as well. To facilitate obfuscation throughout the circuit, \sol{} injects \rx{}-gate pairs with random angles throughout the circuit, as shown in Fig.~\ref{fig:overview} (Step \circled{2}).

A pair consists of an \rx{} gate with a randomly chosen angle and its inverse, an \rx{} gate with the negative of the first angle, injected on the same qubit in direct sequence. We form pairs to ensure that whatever computational logic is inserted is immediately reverted and has no impact on the original program semantics -- we only want the \x{}-gate injections at the circuit end to affect the output as we can control its effect by decoding using the key. We choose the \rx{} gate instead of other rotation gates as it provides sufficient diversity in terms of obfuscation because it decomposes into multiple hardware basis gates (\rz{}, \sx{}, and \x{}), allowing for diverse obfuscation patterns. One may ask how inserting pairs of \rx{} gates obfuscates the structure since it does not affect the computation at all. We explain this next, along with how we choose the circuit locations to inject the \rx{} gates.

\begin{figure}[t]
    \centering
    \includegraphics[scale=0.53]{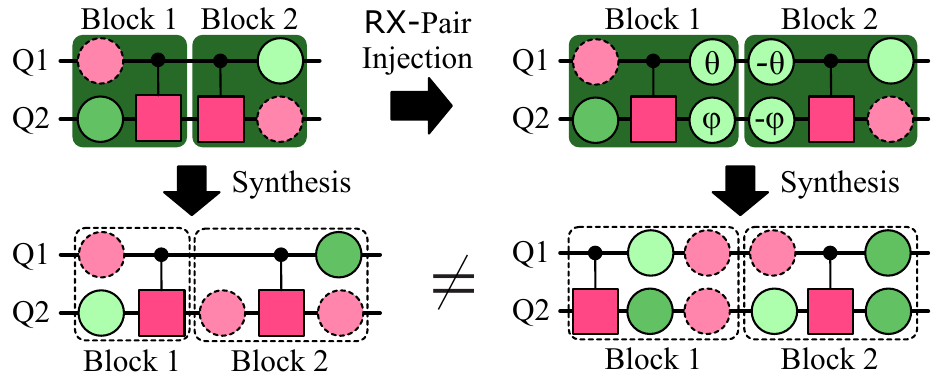}
    \vspace{1mm}
    \hrule
    \vspace{-3mm}
    \caption{\rx{}-pair injection obfuscates the circuit by generating synthesized circuits with different structures.}
    \vspace{-4mm}
    \label{fig:rx_pair}
\end{figure}

\subsection{Circuit Block Formation for Synthesis}

We want to be able to hide the injected \rx{} gates by generating new circuit logic. The way to achieve this is by using the process of synthesis. Synthesis takes a unitary matrix $U$ corresponding to a circuit $C$ and generates a new circuit $\hat{C}$ such that $\hat{C}$'s unitary representation is also $U$. However, this procedure scales exponentially in the number of qubits as the unitary dimensions are $2^n\times2^n$ for an $n$-qubit circuit. Therefore, to perform this operation in a scalable manner, we must divide the circuit into manageable blocks such that each block can be separately synthesized and the synthesized blocks can then be put back together to form the full quantum circuit. \sol{} restricts the size of the blocks to two qubits as the small size allows for scalable and exact synthesis~\cite{khaneja2001cartan} --- the synthesis of bigger sized blocks may be accelerated by generating approximate solutions, however, we want to restrict our synthesis to exact solutions so as to not add unnecessary error to the original program output.

Fig.~\ref{fig:overview} (Step \circled{3}) provides a visual example of how these blocks are formed. In the example, we form three two-qubit blocks. Note that the injected \x{} gates at the end can be absorbed into the last-most block for their respective qubits. The example also shows how \textit{the \rx{} gates are injected at the edges of the blocks. This allows for one gate from each pair to get absorbed into two separate blocks.} This means that each block now represents a different quantum logic than it would in the original circuit without the \rx{}-gate injection. For example, the unitary representing block 3 would be different if it did not have the $\rx{}(-\theta{})$ gate on qubit 2 and $\rx{}(-\phi{})$ gate on qubit 3. When these blocks are now synthesized, they will generate very different logic compared to if they did not have the \rx{} gates (depicted in Fig.~\ref{fig:rx_pair}). Thus, if one were to remove any given block or a region consisting of multiple blocks from the circuit, it obfuscates the logic. In fact, even if they executed that region, the circuit will generate meaningless output as the logic of the sub-regions is altered by \rx{} gate injections. As a note, there is no benefit to having full pairs within a block as the computation cancels itself and will have no impact on the synthesis of that block. Note also that these \rx{}-gate injections have no impact on the overall circuit output, even though they scramble the structure and output of all the blocks and sub-regions.

 \begin{algorithm}[t]
 \caption{Algorithm for dividing the circuit into blocks.}
 \label{alg:blocks}
 {\small
 \begin{algorithmic}[1]
 \State $G \gets$ All of the circuit gates in topological order
 \State $B \gets \{\emptyset{}\}$ \Comment{Set of all blocks, each block is a set of gates} 
 \State \textbf{for} $g$ \textbf{in} $G$ \textbf{do}
 \State \hspace{3mm} \textbf{if} $g$ is a one-qubit gate \textbf{then}
 \State \hspace{7mm} $q \gets$ The qubit that $g$ runs on
 \State \hspace{7mm} \textbf{if} $b$ includes $q$ for any $b \in B$ \textbf{then} add $g$ to $b$
 \State \hspace{7mm} \textbf{else} create a new $k \in B$ and add $g$ to $k$
 \State \hspace{3mm} \textbf{else if} $g$ is a two-qubit gate \textbf{then}
 \State \hspace{7mm} $q1, q2 \gets$ The two qubits that $g$ runs on
 \State \hspace{7mm} \textbf{if} $b$ includes $q1, q2$ for any $b \in B$ \textbf{then} add $g$ to $b$
 \State \hspace{7mm} \textbf{else if} $b$ includes one of $q1, q2$ for any $b \in B$ \textbf{then}
 \State \hspace{11mm} Complete block $b$, create a $k \in B$, and add $g$ to $k$
 \State \hspace{7mm} \textbf{else} create a new $k \in B$ and add $g$ to $k$
 \State \Return $B$
 \end{algorithmic}}
 \end{algorithm}

\begin{figure}[t]
    \centering
    \vspace{-4mm}
    \includegraphics[scale=0.45]{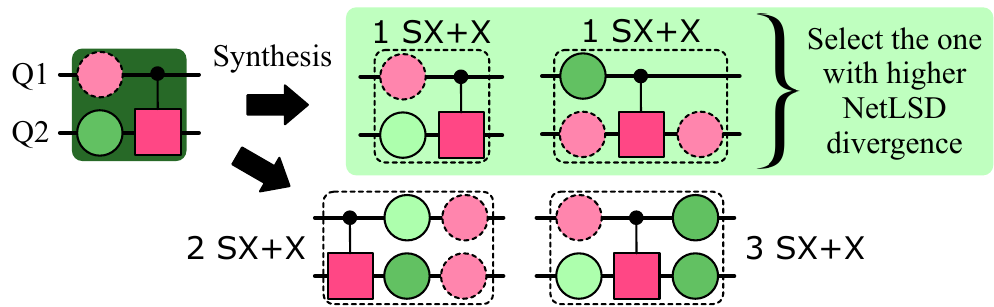}
    \vspace{1mm}
    \hrule
    \vspace{-3mm}
    \caption{\sol{} selects the synthesized circuit with the lowest \sx{}+\x{} count and the highest structural difference as compared to the original block circuit.}
    \vspace{-4mm}
    \label{fig:synth}
\end{figure}

Algorithm~\ref{alg:blocks} shows the pseudo-code for how these blocks are formed for a quantum circuit. The circuit gates are processed one by one in topological order -- which is the order of the gates from the left to the right of the circuit -- and assigned to different blocks. A block is continued to be formed until a two-qubit gate is encountered that has one qubit in one block and the second qubit in another block or the circuit end is reached. The complexity of this algorithm is $O(g)$, where $g$ is the number of gates in the circuit as it has to go through each gate only once. While there are other methods of forming these blocks (e.g., greedy heuristics to form deeper blocks), we chose this method due to its speed and result quality (Sec.~\ref{sec:evaluation}). Once the blocks are formed, the \rx{} gates are injected at the edges. Note that an adversary cannot readily identify their edges and injection locations as the logic gets altered by synthesis and the block boundaries do not persist in the final full circuit.

Next, we describe the synthesis procedure. 

\vspace{-2mm}

\subsection{Block Synthesis to Generate the Final Circuit}

As mentioned earlier, there are a large number of different ways of realizing the same quantum logic. That means that a given unitary matrix, $U$, can be realized using many different circuit structures. We use synthesis to take a unitary matrix $U$ corresponding to a block circuit $C$ and generate a new circuit $\hat{C}$ such that $\hat{C}$'s unitary representation is also $U$. The $U$ matrix for a block is simply calculated by multiplying the matrices (Kronecker product) of all the gates within the block. While there are many different ways of synthesizing a $U$, we use KAK decomposition~\cite{khaneja2001cartan} to construct the new circuit gate by gate, as it is an efficient method to exactly synthesize two-qubit unitaries. \textit{We generate $k$ circuits for the same block such that the generated circuits have the same number of two-qubit gates as the original block circuit because two-qubit gates tend to be highly noisy and can degrade the output quality.} Note that just because the synthesized circuit has the same number of two-qubit gates as the original circuit, does not imply that those gates will be in the same position and orientation, which adds to the obfuscation.

On the other hand, we allow leeway in terms of the number of one-qubit physical (\sx{}$+$\x{}) and virtual (\rz{}) gates as these gates have little-to-none error effects and therefore, can be of help for obfuscation. A na\"ive approach would call for selecting the circuit with the least number of \sx{}$+$\x{} gates out of the $k$ generated circuits as that would have the least noise effects. However, \sol{} also considers the difference in the structure of the generated block and the original block for the purpose of obfuscation. To compare the structures of two circuits, it converts them into Directed Acyclic Graphs (DAGs) and uses the NetLSD divergence~\cite{tsitsulin2018netlsd,banerjee2012structural} metric to assess their degree of similarity. NetLSD is calculated by computing the spectra of the normalized Laplacian matrices corresponding to the two DAGs and comparing their spectral node signature distributions. Thus, out of the generated $k$ circuits, \sol{} takes the top circuits with the fewest \sx{}$+$\x{} gates, and out of these circuits, it selects the one with the highest NetLSD divergence to the original block circuit (Fig.~\ref{fig:synth}). This ensures that \sol{} achieves a balance between gate count minimization and structural difference maximization.

Fig.~\ref{fig:overview} (Step \circled{4}) shows three aspects of note. First, the synthesized blocks have the same number of \cx{} gates as the original blocks, but the gates are at different positions and have different orientations. Second, the synthesized blocks can have more or less one-qubit gates than the original circuit and these gates can be in completely different positions structurally. Generally, the synthesized block has more one-qubit gates than the original block for stronger obfuscation, as we demonstrate in the evaluation (Sec.~\ref{sec:evaluation}). Last, the figure shows the final fully encoded circuit is structurally completely different than the initial (original) circuit, with all traces of \x{}-gate and \rx{}-gate injections and block boundaries erased (as we evaluate in Sec.~\ref{sec:evaluation} using a structural distance metric). If an adversary were to separate any region of the circuit and manually inspect it, they would observe a completely different structure and output than they would have otherwise. In fact, it is not possible to even perform one-to-one correspondence of any region in the original and encoded circuits due to gate injections and synthesis.

Next, we discuss how the user can decode the output.

\begin{figure}[t]
    \centering
    \includegraphics[scale=0.54]{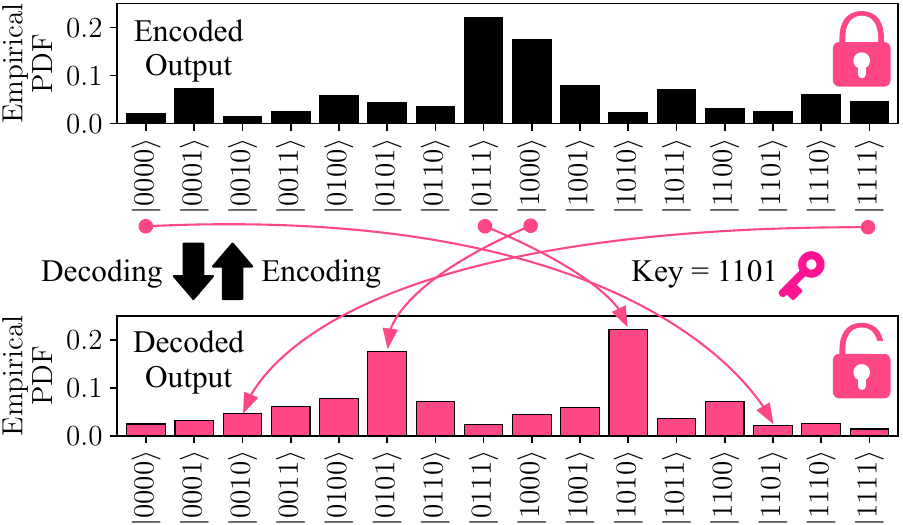}
    \vspace{1mm}
    \hrule
    \vspace{-3mm}
    \caption{An illustration of how the probability distribution gets shifted during the encoding procedure and can be revered back using the key using the decoding procedure.}
    \vspace{-4mm}
    \label{fig:decoding}
\end{figure}

\subsection{Output Recovery Using \sol{} Decoding}

As depicted in Fig.~\ref{fig:decoding}, \sol{}'s decoding is a relatively quick step that is performed on the user side using classical resources. As the user has access to the encoding key, they can simply flip the bits in the output states in accordance with the key. As the figure shows, the user receives the encoded output in which probabilities of all the states are scrambled. If we consider the two states with the highest probabilities, $\ket{0111}$ and $\ket{1000}$, and apply the key $1101$, we get $\ket{0111} \rightarrow \ket{1010}$ and $\ket{1000} \rightarrow \ket{0101}$. Similarly, all the states can be decoded to get the complete output distribution.

That concludes the discussion of \sol{}'s  design. Next, we  present its evaluation methodology. 
\section{\sol{}'s Evaluation Methodology}
\label{sec:methods}

\begin{table}[t]
    \centering
    \caption{The algorithms used to evaluate \sol{}.}
    \vspace{-3mm}
    \scalebox{0.86}{
    \begin{tabular}{|p{17mm}|p{67mm}|}
        \hline
        \textbf{Algorithm} & \textbf{Description} \\
        \hline
        \hline
        \rowcolor[RGB]{200, 255, 200} ADD & Quantum Adder Circuit~\cite{cuccaro2004new} \\
        ADV & Google's Quantum Advantage Algorithm~\cite{ arute2019quantum} \\
        \rowcolor[RGB]{200, 255, 200} DNN & Deep Quantum Neutral Network~\cite{stein2022quclassi} \\
        HLF & Hidden Linear Function Circuit~\cite{bravyi2018quantum} \\
        \rowcolor[RGB]{200, 255, 200} MULT & Quantum Multiplier Circuit~\cite{hancock2019cirq} \\
        QAOA & Quantum Alternating Operator Ansatz~\cite{farhi2016quantum} \\
        \rowcolor[RGB]{200, 255, 200} QFT & Quantum Fourier Transform~\cite{namias1980fractional} \\
        SAT & Quantum Boolean Satisfiability Algorithm~\cite{su2016quantum} \\
        \rowcolor[RGB]{200, 255, 200} TFIM & Transverse Field Ising Model~\cite{bassman2021arqtic} \\
        VQE & Variational Quantum Eigensolver~\cite{mcclean2016theory} \\
        \rowcolor[RGB]{200, 255, 200} WSTATE & W-State Preparation/Assessment Circuit~\cite{fleischhauer2002quantum} \\
        XY & XY Quantum Heisenberg Model Circuit~\cite{bassman2021arqtic} \\
        \hline
    \end{tabular}}
    \vspace{-4mm}
    \label{tab:algorithms}
\end{table}

\noindent\textbf{Experimental Setup.} We evaluate \sol{} using IBM's quantum computing cloud platform~\cite{castelvecchi2017ibm}. This platform provides support for ideal quantum simulation, noisy quantum simulation, and real-hardware execution. To implement and evaluate \sol{} on the client side, we used Qiskit~\cite{aleksandrowiczqiskit} (version 0.36.0), which is a Python-based (version 3.9.7) programming framework that implements a wide range of quantum computing features. We used Qiskit's Aer library (version 0.10.4) to convert blocks into unitary matrices for synthesis. The Qiskit transpiler was used to synthesize blocks into gate logic using KAK decomposition~\cite{khaneja2001cartan}, convert circuits into IBM-compatible basis gates, and run optimization passes during compilation. We set the number of circuits generated for each block to select from to be 3 ($k = 3$) as we observe diminishing returns beyond that point. Qiskit converters were used to convert circuits into NetworkX~\cite{hagberg2020networkx} (version 2.6.3) graph representations, and the NetRD~\cite{mccabe2020netrd} library (version 0.3.0) was used to calculate distances between two graphs to assess the similarity of two circuits. All of the above steps were run entirely on the client side. To run quantum simulations and real-hardware executions on the IBM cloud, we used Qiskit's IBMQ Provider library (version 0.19.0).

\vspace{2mm}

\noindent\textbf{Algorithms and Benchmarks.} Table~\ref{tab:algorithms} lists the benchmarks used to evaluate \sol{}. The evaluated benchmarks represent a wide variety of quantum algorithms in terms of algorithmic domain, circuit structure, and circuit size. For example, TFIM and XY are Hamiltonian evolution algorithms, while QAOA and VQE are optimization algorithms. The circuit size varies from 4 qubits to 128 qubits.

\vspace{2mm}

\noindent\textbf{Quantum Cloud Platform.} We used the IBM cloud as the quantum hardware provider to run several types of experiments. We used IBM's QASM simulator to run ideal simulations (i.e., no noise) to produce the ideal circuit outputs. We ran $100,000$ shots per circuit. This is the number of times a circuit is prepared, run, and measured to generate the output distribution. A circuit has to be run multiple times because each run only produces one output state (bit string). We simulate circuits up to 32 qubits in size, as it is not possible to simulate circuits beyond that size due to the exponential scaling requirements of simulating quantum algorithms. For larger algorithms, we show the circuit characteristics as they can be compiled and synthesized, but not run or simulated. 

We used three different IBM quantum computers for real-hardware algorithm executions; the computer was chosen based on the algorithm size. We used the 127-qubit Washington quantum computer (the largest available) to run circuits larger than 27 qubits. It was not possible to run the 128-qubit circuit as computers of that size are not yet available.  We used IBM's 27-qubit Toronto quantum computer to run circuits of size 8-27 qubits. We used IBM's seven-qubit Lagos quantum computer to perform real-hardware executions for circuits of size seven qubits or less. We ran $32,000$ shots per circuit, as it is the maximum allowed. We also used this computer to perform an in-depth case study using the QAOA variational algorithm.  The four hardware-compatible basis gates on the IBM computers are \cx{}, \sx{}, \x{}, and \rz{}. Therefore, all of our analysis is with respect to these gates. However, \sol{} is compatible with any other basis gate set.  

\vspace{2mm}

\noindent\textbf{Evaluation Metrics.} To assess the quality of the output, we use two metrics. (1) The \textbf{Total Variation Distance (TVD)} is used to quantify the difference between two output probability distributions. The TVD of two outputs is calculated as $\frac{1}{2}\sum_{k=1}^{2^n}\abs{p_1(k) - p_2(k)}$, where $p_1(k)$ is the probability of state $k$ in the first output and $p_2(k)$ is the probability of state $k$ in the second output. The closer the TVD is to 1, the bigger the difference between the two output distributions; the closer it is to 0, the smaller the difference. The TVD is used to compare \sol{}'s output to the baseline circuit's output and the ideal output. (2) The \textbf{Dominant State Percentile} is used to quantify how close a technique is to identifying the true dominant or winner solution state, i.e., the state with the highest probability. Many quantum algorithms have one solution that needs to be identified. If that state is the one with the highest probability, then it is in the 100$^{\text{th}}$ percentile. The lower the percentile, the more difficult it is to identify the dominant state. However, a value of less than 100$^{\text{th}}$ percentile, no matter how close it is to the 100$^{\text{th}}$ percentile, indicates that the technique has failed to identify the dominant state.

To assess the structural similarities of any two circuits, we convert them into Directed Acyclic Graphs (DAGs) and use a widely-used graph distance metric, NetLSD, to examine their differences. The \textbf{NetLSD divergence}~\cite{tsitsulin2018netlsd,banerjee2012structural} is calculated by computing the spectra of the normalized Laplacian matrices corresponding to the two DAGs and comparing their spectral node signature distributions, i.e., the heat signatures. A value greater than 10$^{2}$ indicates reasonably high dissimilarity~\cite{tsitsulin2018netlsd}.

Lastly, we also look at circuit structure properties such as the \textbf{number of \cx{} gates} (two-qubit gates), \textbf{the number of \sx{}+\x{} gates} (one-qubit physical gates), and \textbf{the number of \rz{} gates} (one-qubit virtual gates). These metrics help us analyze how much the circuit size increases due to \sol{}'s obfuscation process. We also look at the \textbf{circuit depth}, which is the number of \cx{} gates in the critical path of the circuit.

For the QAOA case study, we also assess the \textbf{expectation loss}, which is the average loss incurred in each iteration of optimizing the parameters of the variational circuit, and visually investigate the structural characteristics of different circuits. Further details about the methodology of this case study are included alongside its evaluation (Sec.~\ref{sec:evaluation}).

\vspace{2mm}

\noindent\textbf{Competitive Techniques.}  As \sol{} is the first technique to implement obfuscation of quantum circuit outputs, there are no competitive techniques in this domain. Therefore, we compare \sol{} to the \textbf{baseline technique}, which performs \textit{no output or structural obfuscation}, but benefits from all the optimizations supported by the Qiskit compiler, including synthesis for gate count reduction. This allows us to compare \sol{} w.r.t. circuit characteristics (e.g., increase in the number of quantum gates and structure of the circuit) and solution quality (e.g., identification of dominant solution state, program output error in terms of TVD). 

Additionally, we compare \sol{} to the \textbf{uncorrected} output, i.e., when \sol{}'s decoder is not deployed, to demonstrate the strength of \sol{}'s obfuscation. We also investigate the strength of the two obfuscation techniques employed by \sol{} on their own by comparing to versions of \sol{} that only implement \textbf{circuit-end \x{}-gate injections} or only implement \textbf{throughout-circuit \rx{}-gate injections}.
\section{\sol{}'s Evaluation and Analysis}
\label{sec:evaluation}

\begin{figure}[t]
    \centering
    \includegraphics[scale=0.55]{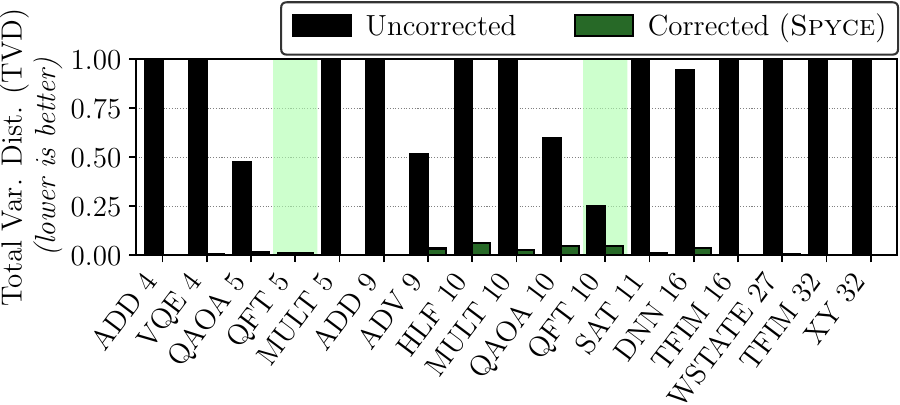}
    \vspace{1mm}
    \hrule
    \vspace{-3mm}
    \caption{When \sol{}'s output is not corrected, it has a high TVD relative to the baseline output. When it is corrected using \sol{}'s decoder, the TVD becomes negligible. QFT is the only exception due to its uniform output distribution. The numbers next to the algorithm names indicate the corresponding number of qubits. }
    \vspace{-5mm}
    \label{fig:tvd}
\end{figure}

\begin{figure}[t]
    \centering
    \includegraphics[scale=0.55]{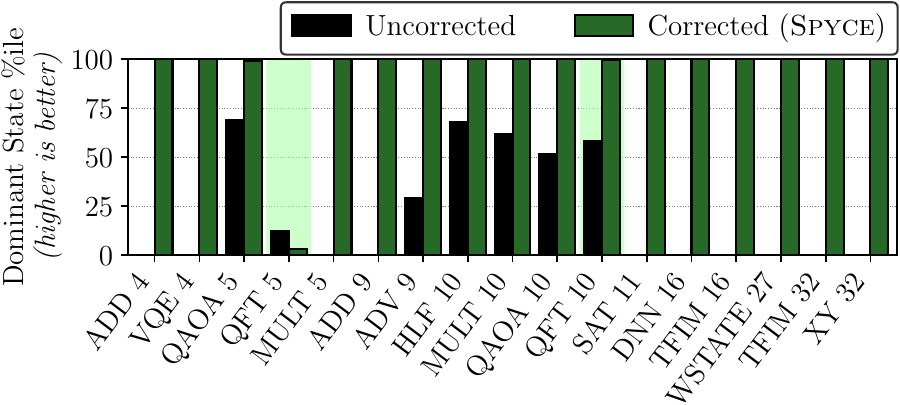}
    \vspace{1mm}
    \hrule
    \vspace{-3mm}
    \caption{For algorithms with a dominant output state (all but QFT), the uncorrected output cannot identify it for any of the algorithms, while the output corrected with the \sol{} decoder can identify the dominant state for all algorithms.}
    \vspace{-4mm}
    \label{fig:dom}
\end{figure}

\begin{figure*}
    \centering
    \includegraphics[scale=0.58]{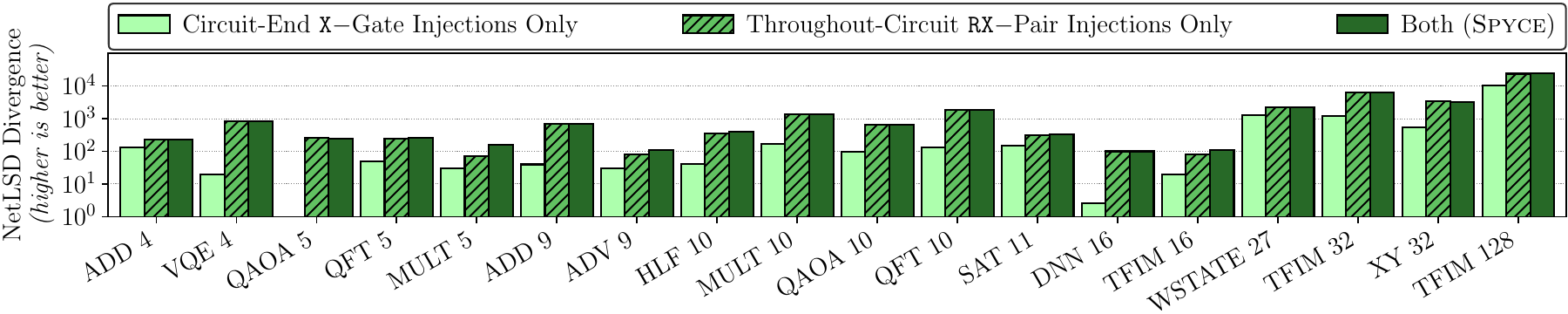}
    \vspace{1mm}
    \hrule
    \vspace{-3mm}
    \caption{The comparison of the NetLSD divergence of the original circuit with the corresponding \sol{}-generated circuits shows that \rx{}-pair injections help generate circuits that are structurally very different than the original circuit. On the other hand, the circuit-end \x{}-gate injections help obfuscate the output but not the structure. Note the log scale.}
    \vspace{-4mm}
    \label{fig:distance}
\end{figure*}

\begin{figure*}
    \centering
    \includegraphics[scale=0.58]{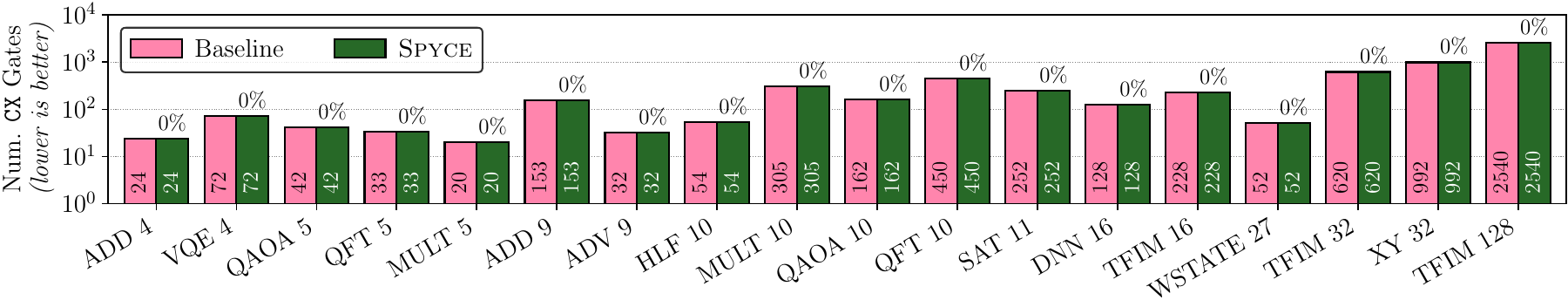}
    \vspace{1mm}
    \hrule
    \vspace{-3mm}
    \caption{\sol{} ensures that there is no increase in the number of \cx{} gates in the synthesized encrypted circuits compared to the original circuits across all algorithms as these are high-error operations, which can cause substantial output errors. The numbers inside the bars indicate the raw gate counts for the respective techniques and the numbers atop \sol{}'s bars indicate the percent increase in \sol{}'s counts relative to the baseline. Note the log scale.}
    \vspace{-4mm}
    \label{fig:num_cx}
\end{figure*}

\begin{figure*}
    \centering
    \includegraphics[scale=0.58]{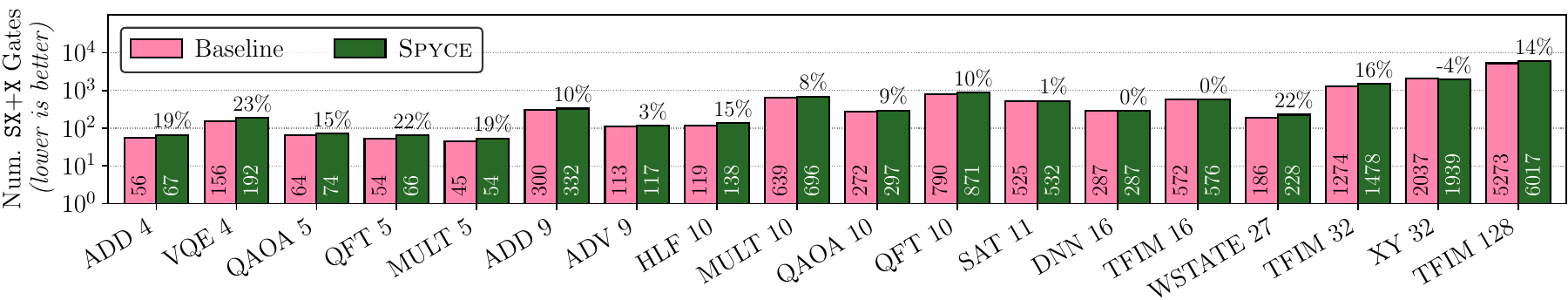}
    \vspace{1mm}
    \hrule
    \vspace{-3mm}
    \caption{\sol{} increases the number of one-qubit physical gates (\sx{} and \x{}) by 13.6\% on average for the purpose of structural and output obfuscation. It chooses to increase these gates as they have negligible noise effects as compared to \cx{} gates.}
    \vspace{-4mm}
    \label{fig:num_sx}
\end{figure*}


\subsection{\sol{}'s Obfuscation Effectiveness}

We first analyze \sol{}'s effectiveness to obfuscate the program output of the quantum circuits. Fig.~\ref{fig:tvd} shows the total variation distance (TVD) between the output produced by the baseline circuit and the output produced by \sol{}'s obfuscated circuit, but not corrected by the \sol{} decoder. The bars representing this TVD are labeled as ``Uncorrected''. 

Recall that the baseline circuit does not employ any obfuscation. If \sol{} is completely ineffective, then the  \sol{} without the decoder (``Uncorrected'') bars would result in almost zero TVD -- that is, despite the obfuscation, one does not need to decode the output as the uncorrected output is already the same as the baseline circuit without obfuscation. However, we observed, this is indeed not the case. In most cases (Fig.~\ref{fig:tvd}), the TVD bars for the ``Uncorrected'' case is almost close to one, as desired.   

Next, we observe that Fig.~\ref{fig:tvd} shows another set of bars after the \sol{} decoder is applied, and the output is corrected on the client side(labeled as ``Corrected''). Once the \sol{} decoder is applied, the TVD between \sol{}'s output and the baseline output is reduced to a negligible amount ($<$0.05 across all algorithms). This indicates that the corrected \sol{} output is very similar to the original output, demonstrating the effectiveness of the \sol{}. As a note, two QFT circuits of different sizes show low TVD for the ``Uncorrected'' case also. This is because the QFT program scrambles the input and generates an equal probability across all output states. In this case, even if \sol{} scrambles and shifts the probabilities of the output states, all states still have similar probabilities. 

While the TVD results are promising, another important metric is the ability to identify the dominant state in the output of the quantum algorithm. Fig.~\ref{fig:dom} shows the percentile at which the actual dominant state (i.e., the dominant state in the output of the baseline circuit) falls in the uncorrected output as well as the corrected output. Ideally, this should be 100\%. The uncorrected output is not able to identify the dominant state for any of the algorithms. In fact, across all algorithms, the dominant state falls in the 75$^{th}$ percentile or less. For most algorithms, it is in the 0$^{th}$ percentile. This means that the dominant state is not even close to being the highest probability state in the uncorrected output for any of the algorithms. Thus, an adversary, with access to the obfuscated output, cannot identify the dominant state of the algorithm without the key to decode the output.

On the other hand, when the output is corrected by the \sol{} decoder, the dominant state is identified by \sol{} across all algorithms with a dominant state (which are all algorithms except for QFT). Recall that in the case of QFT, while there is no dominant state in the ideal scenario because the simulation has statistical variability and generates one state with a slightly higher probability than others in the baseline circuit output, the results simply indicate the ability of \sol{} to match that state. 

Overall, \textit{\sol{} is successfully able to obfuscate the output probability distribution and the dominant state from an adversary. Moreover, it is also successfully able to recover locally on the user's end when the output is returned by the hardware provider and decoded using the key.} Next, we look at how effective \sol{} is at obfuscating the circuit structure.

\subsection{\sol{}'s Impact on Circuit Structure}

The purpose of this evaluation is to show that \sol{} produces a significantly different circuit structure than the original circuit and the need for injecting \rz{}-gate pair injection.

Fig.~\ref{fig:distance} shows the NetLSD divergence metric to compare the structures of the \sol{}-generated circuits to the baseline circuits. The figure also shows the distance of the baseline circuit to the circuit with only synthesized \x{}-gate injections at the end, the circuit with only synthesized \rz{}-gate injections throughout the circuit, and the combined effect (\sol{}).

The circuit-end \x{}-gate injections obfuscate the output effectively, but we observe that the circuit-end \x{}-gate injections has limited impact on obfuscating the circuit structure (Fig.~\ref{fig:distance}). This is because synthesizing the \x{} gates into the circuit only modifies the circuit structure at the end of the circuit, but the rest of the circuit structure would remain the same as the baseline (original) circuit. Thus, while \x{} gate injection is an effective technique to obfuscate the output, it is not sufficient to obfuscate the circuit structure -- motivating our design element to inject \rx{} gates throughout the circuit.

Indeed, injecting the \rx{} gates throughout the circuit, dividing them into disjoint blocks, and synthesizing the blocks shows promising results. Recall that injecting only the \rx{} gates does obfuscate the output; \rx{} gates need to be injected in pairs to ensure that the program intent is not modified. 

Fig.~\ref{fig:distance} shows that for all of the algorithms, the structural obfuscation that injecting the \rx{} gates provides is orders of magnitudes greater than structural obfuscation facilitated by the circuit-end \x{}-gate injections. As an instance, for the MULT 10 algorithm, the NetLSD divergence with only circuit-end \x{}-gate injections is $1.7\times{}10^{2}$, with only throughout-circuit \rx{}-gate injections is $1.4\times{}10^{3}$, and with both (\sol{}) is also $1.4\times{}10^{3}$. The trend is similar for other algorithms; they all have a divergence of $>10^{2}$ with \sol{}.

\textit{These results confirm that the combination of \x{}-gate and  \rx{}-pair injection is required and effective to achieve both desired outcomes: output obfuscation and structural obfuscation}. Next, we study \sol{}'s impact on gate count and circuit depth.

\begin{figure}[t]
    \centering
    \includegraphics[scale=0.55]{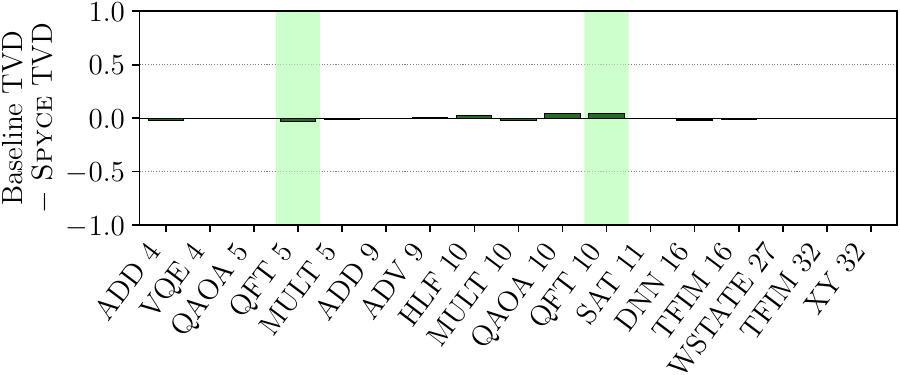}
    \vspace{1mm}
    \hrule
    \vspace{-3mm}
    \caption{Execution on real quantum hardware shows that \sol{}'s corrected output is able to achieve a similar TVD to the ideal output as the baseline output.}
    \vspace{-3mm}
    \label{fig:real_tvd}
\end{figure}

\begin{figure}[t]
    \centering
    \includegraphics[scale=0.55]{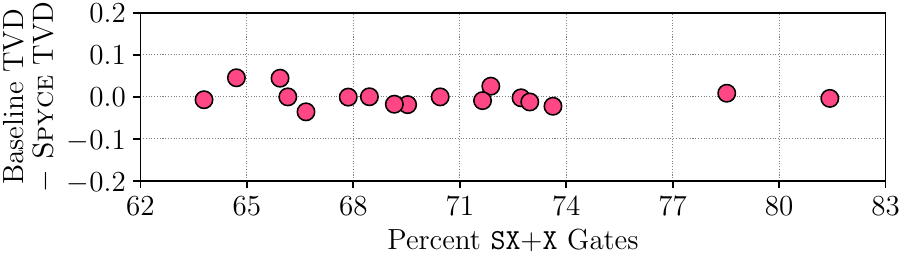}
    \vspace{1mm}
    \hrule
    \vspace{-3mm}
    \caption{The TVD difference between the baseline and \sol{} outputs does not increase with increase in the percent of \sx{}+\x{} gates in the circuit (percent relative to \sx{}+\x{}+\cx{}).}
    \vspace{-3mm}
    \label{fig:sx_impact}
\end{figure}

\begin{table}[t]
    \centering
    \caption{Compilation times of the algorithms in seconds.}
    \vspace{-3mm}
    \scalebox{0.8}{
    \begin{tabular}{|>{\columncolor[RGB]{200, 255, 200}}p{17mm}|p{13mm}|p{10mm}||>{\columncolor[RGB]{200, 255, 200}}p{17mm}|p{13mm}|p{10mm}|}
        \hline
        \textbf{Algorithm} & \textbf{Baseline} & \textbf{\sol{}} & \textbf{Algorithm} & \textbf{Baseline} & \textbf{\sol{}} \\
        \hline
        \hline
        ADD 4 & 2.4 & 4.1 & QAOA 10 & 6.3 & 12.0 \\
        \hline
        VQE 4 & 4.0 & 8.0 & QFT 10 & 15.0 & 32.8 \\
        \hline
        QAOA 5 & 2.4 & 4.0 & SAT 11 & 8.6 & 22.9 \\
        \hline
        QFT 5 & 2.4 & 3.9 & DNN 16 & 6.1 & 11.4 \\
        \hline
        MULT 5 & 2.2 & 3.2 & TFIM 16 & 10.0 & 20.6 \\
        \hline
        ADD 9 & 6.9 & 13.5 & WSTATE 27 & 5.3 & 10.3 \\
        \hline
        ADV 9 & 3.3 & 8.0 & TFIM 32 & 27.6 & 54.9 \\
        \hline
        HLF 10 & 3.7 & 7.0 & XY 32 & 33.3 & 93.2 \\
        \hline
        MULT 10 & 11.7 & 24.1 & TFIM 128 & 80.0 & 230.4 \\
        \hline
    \end{tabular}}
    \vspace{-3mm}
    \label{tab:compile_times}
\end{table}

\subsection{\sol{}'s Impact on Gate Count and Circuit Depth}

\sol{} injects additional different types of gates to achieve program output obfuscation. Unfortunately, such injections can increase the gate count and circuit depth, which can have side effects on the solution quality produced by \sol{} on real noisy quantum hardware (i.e., \sol{}'s circuit producing higher TVD than the original unobfuscated circuit). 

First, we compare the number of \cx{} gates in the \sol{}-generated circuits to the number of \cx{} gates in the baseline circuits in Fig.~\ref{fig:num_cx}. The figure shows that \textit{across all the algorithms, the number of \cx{} gates remains the exact same as the baseline circuit}. This is by design, as \sol{} ensures during the block synthesis process that the number of \cx{} gates are not increased to avoid the impact of their high error rates on the program output on noisy quantum computers. This enables \sol{} to not increase the output error, as we see in the next subsection. Note also that because of the \cx{} count remaining the same and the synthesis procedure only being applied to two-qubit blocks, \sol{} also does not incur any increase in circuit depth (number of \cx{} gates on the critical path). We do not show this result to avoid repetition as all algorithms incur a 0\% increase in circuit depth.

Second, we compare the number of \sx{}+\x{} gates in the \sol{}-generated circuits to the number of \sx{}+\x{} gates in the baseline circuits in Fig.~\ref{fig:num_sx}. Even though these two one-qubit gates are different in terms of their impact on the quantum state, we group the two together because they have the same error rates as far as the hardware noise effects are concerned. On average, the number of \sx{}+\x{} gates is increased by 13.6\%, and the number of \rz{} gates is increased by 11.6\% (not shown for brevity; \rz{} gates are virtual gates and have no contribution to the error effects). While the increases are non-negligible, they do not have much impact on the output noise -- as we will observe in the next  part of the evaluation where we demonstrate real-hardware results -- due to the little-to-none error impact of the one-qubit gates.

\subsection{\sol{}'s Effectiveness on Real NISQ Hardware and its Compilation Overhead}

We now examine \sol{}'s performance on real quantum computers for algorithms up to 32 qubits. The noise levels only impact the output and not the circuit structure, therefore, we only analyze the output metrics.

Fig.~\ref{fig:real_tvd} shows the difference between the TVD of the baseline output on noisy IBM computers (relative to the ideal simulation output) and the TVD of \sol{}'s output on noisy IBM computers (also relative to the ideal simulation output). When this metric is positive it indicates that the baseline has a higher error and when it is negative it indicates that \sol{} has a lower error. The figure shows that both TVDs are largely similar (maximum difference of 0.05 or 5\%), i.e., both techniques are similarly impacted by the hardware noise -- this is primarily due to \sol{}'s design element that enforces the \cx{} gate count in the obfuscated circuit  to be the same as the baseline circuit. In fact, as Fig.~\ref{fig:sx_impact} shows, the difference between the baseline's TVD and \sol{}'s TVD remains low even if \sx{}+\x{} gates form a large percent ($>75\%$) of all the physical gates in the circuit (each point in the figure corresponds to one real-hardware-evaluated algorithm). Thus, the additional one-qubit physical gates added by \sol{} do not have a major impact on the output error in noisy environments, as the errors are largely dominated by \cx{} gates.

We also note that due to the high noise level, the baseline output is also not able to identify the dominant state correctly in many cases. Nonetheless, \sol{} is able to identify the dominant state whenever the baseline does (not shown in a figure). These results demonstrate how \sol{} successfully scrambles the output and decodes it in a high-noise setting.

Next, we study the compilation time overhead of \sol{}. Analytically examining the compilation overhead and complexity of \sol{} is challenging, as the overhead of synthesis can vary from algorithm to algorithm based on the intricacies of the circuit structure. Nonetheless,  we empirically analyze the compilation overhead of \sol{}. Table~\ref{tab:compile_times} shows the compilation times of all the algorithms for the baseline circuit and \sol{}-generated circuit. The table shows that the compilation times with \sol{} increase by $2\times{}$ on average over the baseline compilation times. These increases are negligible compared to the queue wait times on quantum cloud, which tend to be in the order of hours~\cite{ravi2021quantum}. \sol{} is, therefore, low-overhead as it obfuscates large circuits for modest increases in compilation times.

\begin{figure}[t]
    \centering
    \includegraphics[scale=0.355]{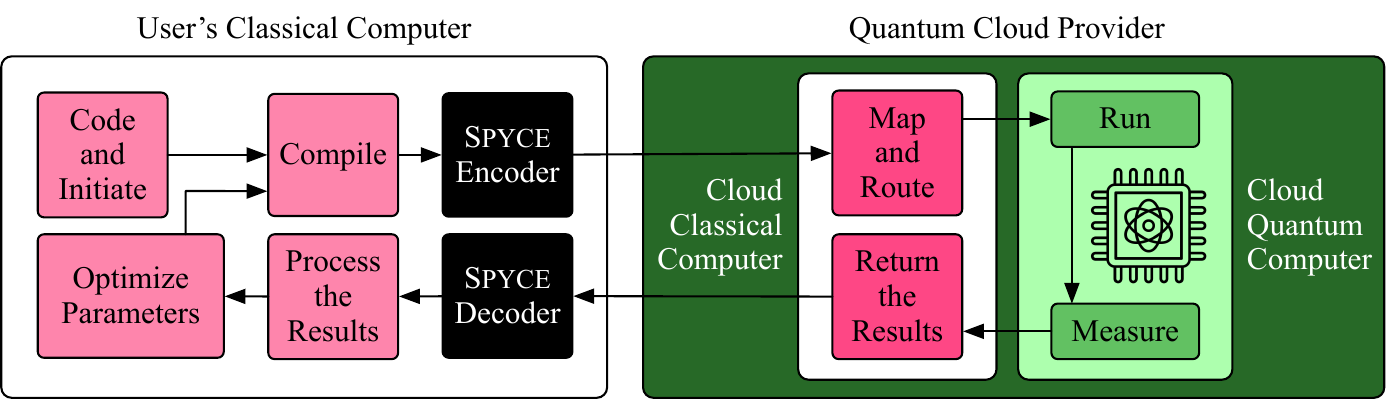}
    \vspace{1mm}
    \hrule
    \vspace{-3mm}
    \caption{Variational quantum algorithms like QAOA follow a hybrid quantum-classical approach to iteratively optimize the parameters of a circuit to achieve a specific objective. \sol{} works to obfuscate each iteration.}
    \vspace{-4mm}
    \label{fig:qaoa_workflow}
\end{figure}

\begin{figure}[t]
    \centering
    \includegraphics[scale=0.32]{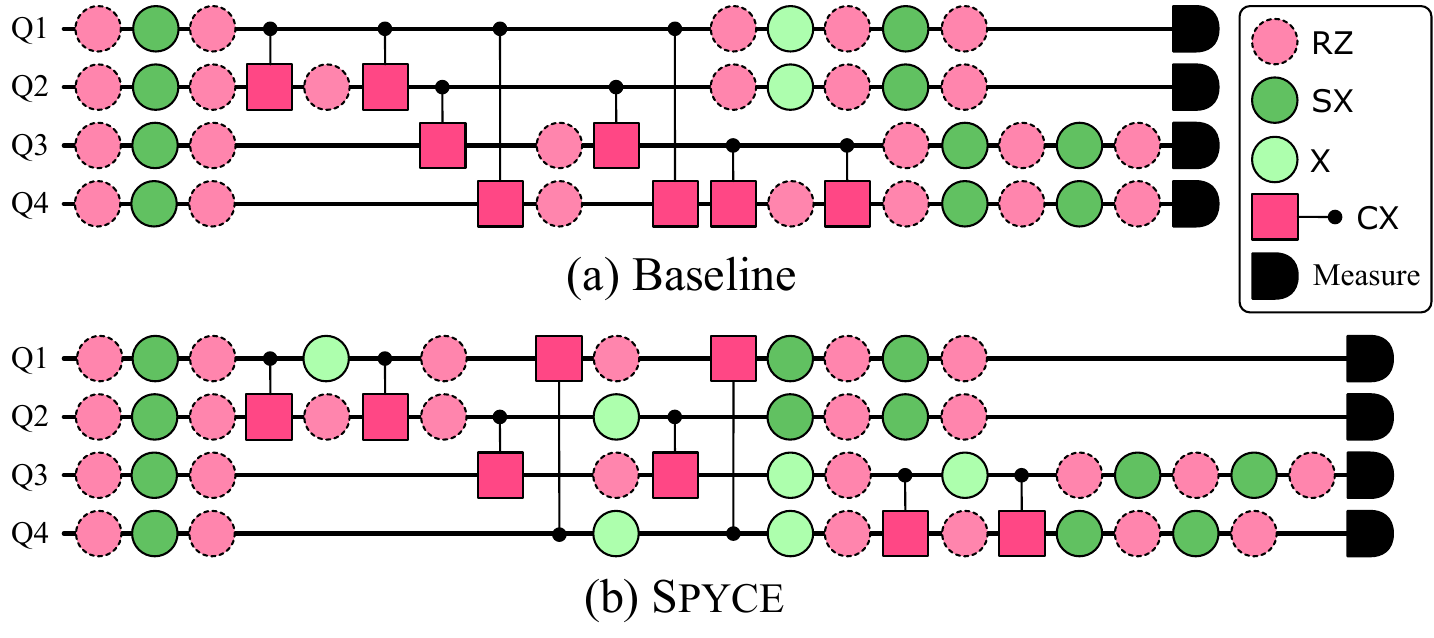}
    \vspace{1mm}
    \hrule
    \vspace{-3mm}
    \caption{Visual circuit structure comparison of the baseline circuit and the \sol{}-generated circuit for the last iteration shows how different the two circuits are with all traces of \x{}- and \rx{}-gate injections being erased due to synthesis.}
    \vspace{-3mm}
    \label{fig:qaoa_circ}
\end{figure}

\begin{figure}[t]
    \centering
    \includegraphics[scale=0.53]{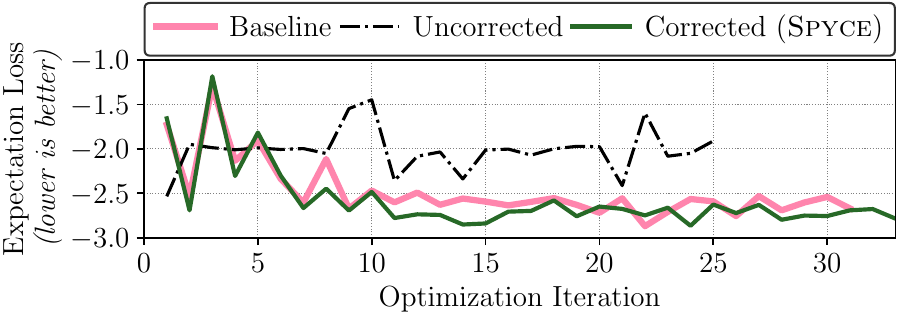}
    \vspace{1mm}
    \hrule
    \vspace{-3mm}
    \caption{The loss curve with \sol{}'s corrected output traces the baseline loss curve. However, when the output is not corrected, the variational algorithm cannot be optimized due to the scrambled results at each iteration.}
    \vspace{-4mm}
    \label{fig:qaoa_loss}
\end{figure}

\begin{figure*}
    \centering
    \includegraphics[scale=0.58]{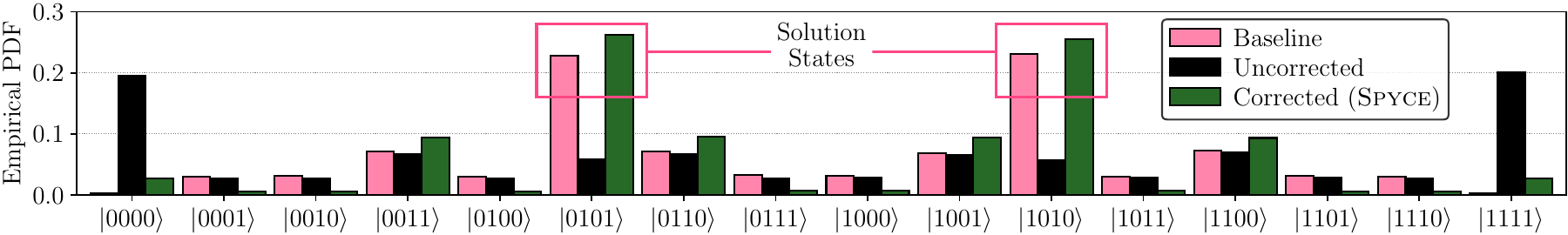}
    \vspace{1mm}
    \hrule
    \vspace{-3mm}
    \caption{The output corrected with \sol{}'s decoder can achieve an output distribution similar to the original circuit's baseline output. It is thus able to identify the two solution states for the QAOA MaxCut problem with the dominant probabilities. In contrast, the obfuscated uncorrected output is not able to identify these states.}
    \vspace{-3mm}
    \label{fig:qaoa_pmf}
\end{figure*}

\subsection{Case Study: Iterative Optimization of a Variational Algorithm using \sol{}}

We now perform a case study of a variational algorithm using the 4-qubit QAOA MaxCut problem. In a variational algorithm, the circuit gate angles are treated as parameters and these parameters are tuned to optimize for a specific objective. Thus, many machine learning or optimization problems fall under the domain of variational algorithms~\cite{mcclean2016theory}. We perform this case study for a variational algorithm because it involves iteratively running the same parameterized circuit with different parameters. Thus, it can be argued that the optimizer can converge to a sub-optimal solution or take longer to converge if \sol{} obfuscates the circuit for each iteration. Our experimental results demonstrate that \sol{} continues to be effective for iterative variational algorithm -- the optimization quality and time to convergence is not impacted by \sol{}'s obfuscation.

Fig.~\ref{fig:qaoa_workflow} shows the iterative structure that a variational algorithm follows with \sol{}. Before shipping out the quantum circuit to the cloud for execution at each iteration, \sol{} decodes the resulting output from the previous iteration before the output is fed to the optimizer, and then, \sol{} obfuscates the circuit in preparation for the next round of optimization. We focus on the 4-qubit QAOA MaxCut problem, as QAOA provides a specific circuit template suitable for many variational problems. The MaxCut problem involves dividing the nodes of a graph into two sets such that the number of edges between the two sets is maximized.

In our case, the graph has four nodes connected to form a rectangle. Therefore, the problem has two solutions ``1010'' and ``0101'', where 0 and 1 are labels for the two sets and the bitstring indicates the set that the nodes belong to in order from 1 to 4. The goal of the optimizer is to adjust the parameters such that these two solutions have the highest probabilities and can be successfully identified. We choose this small problem so that we can investigate it visually.

We first visually examine how the \sol{} generated circuit differs from the baseline circuit in this case. Fig.~\ref{fig:qaoa_circ} shows the two circuits during the last optimization iteration. Note also that the circuit structure for each iteration varies because the synthesis step generates different circuit structures (the parameterized angles vary from one iteration to another). Thus, there is no concern of an adversary being able to infer information about the algorithm -- the concern would be present if the circuit structure remains the same throughout iterations. In the figure, we only show the last-iteration circuits for brevity. The two circuits look substantially different (NetLSD is $2.3\times{}10^2$) in terms of the number of gates and the gate placement within the circuits. In addition, the \rz{} gates in both circuits have very different angles (not shown here for clarity). Thus, \textit{the structure is successfully obfuscated.}

Next, we assess the success of \sol{}'s output obfuscation when run on a real quantum computer (IBM Lagos). Fig.~\ref{fig:qaoa_loss} shows the loss curves over the optimization iterations for three cases: baseline (no obfuscation), uncorrected (\sol{} encoder is applied but the decoder is not, i.e., the perspective of the adversary), and corrected (both \sol{} encoder and decoder are applied). We make several observations.

First, we observe that the loss curve of \sol{} closely follows the loss curve of the baseline circuit. This means that while \sol{} successfully obfuscates every iteration, it is still able to optimize variational algorithms just as efficiently and effectively as the baseline technique -- in other words, the solution quality and time to convergence are not affected. 

Second, when the output is not corrected with the \sol{} decoder but the circuit remains obfuscated, the optimizer is not able to make any progress toward minimizing the loss as each iteration generates scrambled outputs of different types, which cannot be used toward the optimization of a specific objective. This indicates that the cloud provider which has access to only uncorrected output cannot make forward progress in an iterative optimization problem.

Third, Fig.~\ref{fig:qaoa_pmf} shows that \sol{} maintains the solution quality. The corrected \sol{} output distribution over all the states closely resembles the baseline output distribution. Both techniques are able to identify the two solution states as they have the dominant probabilities. On the other hand, the circuit optimized using the uncorrected output has a completely different distribution, where the two states with the highest probabilities are non-solution states. Thus, \textit{an adversary that tries to optimize this problem using the uncorrected output will neither be able to identify the original circuit structure, nor the correct solution. This demonstrates the applicability of \sol{} to variational algorithms.}
\section{Concluding Remarks for \sol{}}
\label{sec:conclusion}

In this work, we introduce \sol{} to protect quantum code and output from adversarial snooping in the quantum cloud. \sol{} leverages circuit-end \x{}-gate injections for output obfuscation and throughout-circuit \rx{}-gate injections for structural obfuscation. Through extensive simulations and real-hardware evaluations, we demonstrate the effectiveness of \sol{} in obfuscating circuit structure and output while maintaining the output fidelity of the algorithm. 

\bibliographystyle{ACM-Reference-Format}
\bibliography{main}

\end{document}